\documentclass[prd,showpacs,showkeys,nofootinbib,twocolumn]{revtex4}

\begin{document}

\title{Motion of test bodies in theories with nonminimal coupling}

\author{Dirk Puetzfeld}
\email{dirk.puetzfeld@aei.mpg.de}
\homepage{http://www.thp.uni-koeln.de/~dp}
\affiliation{Max-Planck-Institute for Gravitational Physics (Albert-Einstein-Institute), Am Muehlenberg 1, 14476 Golm, Germany}

\author{Yuri N. Obukhov}
\email{yo@thp.uni-koeln.de}
\affiliation{Department of Theoretical Physics, Moscow State University, 117234 Moscow, Russia}

\date{ \today}

\begin{abstract}
We derive the equations of motion of test bodies for a theory with nonminimal coupling by means of a multipole method. The propagation equations for pole-dipole particles are worked out for a gravity theory with a very general coupling between the curvature scalar and the matter fields. Our results allow for a systematic comparison with the equations of motion of general relativity and other gravity theories.
\end{abstract}

\pacs{04.25.-g; 04.50.+h; 04.20.Fy; 04.20.Cv}
\keywords{Approximation methods; Equations of motion; Alternative theories of gravity; Variational principles}

\maketitle

%% 04.25.-g Approximation methods; equations of motion
%% 04.50.+h Gravity in more than four dimensions, Kaluza-Klein theory, unified field theories; alternative theories of gravity  
%% 04.20.Fy Canonical formalism, Lagrangians, and variational principles
%% 04.20.Cv Fundamental problems and general formalism

\section{Introduction}\label{introduction_sec}

In a recent work \cite{Bertolami:etal:2007} an alternative gravity theory with nonminimal coupling has been proposed. The theory exhibits -- in contrast to Einstein's theory of gravitation -- a direct coupling between the matter fields and the curvature scalar on the Lagrangian level. In contrast to other theories with extended dynamics in the gravitational sector, such a nonminimal coupling prescription leads to a modification of the equations of motion. The latter were analyzed for specific choices of the matter Lagrangian, e.g., a perfect fluid \cite{Bertolami:etal:2008}.

Here we present a systematic derivation of the equations of motion for arbitrarily structured test bodies. The method employed by us is not tied to a specific choice of the matter Lagrangian and therefore generalizes previous findings. The propagation equations for pole-dipole particles are worked out with the help of a multipole method. In particular it allows us to provide the form of the extra force terms -- ``extra'' in comparison to the case with minimal coupling -- entering the equations of motion. Therefore, our findings should be taken into account in the context of the recent controversy, see \cite{Bertolami:etal:2008,Sotiriou:Faraoni:2008}, regarding the appearance of such an extra term. Furthermore, our results allow for direct a comparison with the motion of test bodies in other gravity theories, in particular the coupling between material and geometrical quantities becomes evident.  

\section{The model under consideration}

In \cite{Bertolami:etal:2007} an extended version of a so-called $f(R)$ gravity theory was considered -- for earlier works on this subject see also \cite{Nojiri:Odintsov:2004,Allemandi:etal:2005}, as well as \cite{Balakin:Kurbanova:2004} for an extension which covers bodies with spin/polarization. Gravity theories in which the usual Einstein-Hilbert Lagrangian is replaced by an arbitrary function of the curvature scalar have attracted a lot of attention during the last few years see, e.g., the reviews \cite{Schmidt:2007,Straumann:2008} and references therein. The $f(R)$-scenario is generalized even further in \cite{Bertolami:etal:2007} by the introduction of a nonminimal coupling term on the Lagrangian level. In particular the following Lagrangian was put forward
\begin{eqnarray}
L_{\rm tot} = \frac{1}{2} f_1\left(R\right) + \left[1+ \lambda f_2\left( R\right) \right] L_{\rm mat}. \label{ansatz_lagrangian}
\end{eqnarray} 
Here $f_1$ and $f_2$ are arbitrary functions of the curvature scalar $R$ and $L_{\rm mat}$ is the matter Lagrangian. The nonminimal coupling of matter and gravity is controlled by the constant $\lambda$. The general field equations -- in terms of the functions $f_1$ and $f_2$ and their derivatives -- are given in \cite{Bertolami:etal:2007}; their explicit form is irrelevant for the subsequent analysis though. Theories of the above mentioned kind, in particular with a nonminimal coupling between the curvature scalar and a scalar field, have also been considered before in a cosmological context see, e.g., the review \cite{Nojiri:Odintsov:2007}.

In contrast to standard $f(R)$ gravity theories, the last term in (\ref{ansatz_lagrangian}) leads to a modification of the equations of motion. As was already shown in (5) of \cite{Bertolami:etal:2007} the usual conservation law -- as, for example, found in general relativity -- is replaced by
\begin{eqnarray}
\nabla^i T_{ij} = \frac{\lambda F_2}{1+\lambda f_2} \left( g_{ij} L_{\rm mat} - T_{ij} \right) \nabla^i R. \label{conservation}
\end{eqnarray}  
Here $F_2\left(R\right):=df_2\left(R\right)/dR $ denotes a shortcut for derivatives of the unspecified function $f_2\left(R\right)$ of the curvature scalar and the energy-momentum tensor of matter is defined in a standard way by $\sqrt{-g}T_{ij} :=  -2\delta (\sqrt{-g}L_{\rm mat})/\delta g^{ij}$.

Of course from (\ref{conservation}) it becomes immediately apparent, that the equations of motion of the theory under consideration differ from the ones of general relativity if the right-hand side (rhs) of (\ref{conservation}) is nonvanishing. In \cite{Bertolami:etal:2007} it is suggested that a deviation of this kind due to the nonminimal coupling term may play a role for the observed flatness of rotation curves \cite{McGaugh:etal:2001,deBlock:etal:2001} or in the context of the so-called Pioneer anomaly \cite{Anderson:etal:2002}. 

In their study of the motion of test bodies, the authors of \cite{Bertolami:etal:2007} made an explicit assumption for the energy-momentum tensor of matter entering (\ref{conservation}), which they assumed to be of perfect fluid form. Also in \cite{Harko:2008} a similar analysis is carried out which is based on a specific choice for the energy-momentum tensor of the system. 

Here we extend the previous analysis by deriving the general form of the equations of motion for test bodies. Our analysis, which relies on a well established multipole approximation technique, is independent of a specific choice of energy-momentum tensor. 

Without going into historical detail we only mention that in the context of general relativity the method -- and variations of it -- was utilized in the works \cite{Mathisson:1931:1,Mathisson:1931:2,Mathisson:1937,Papapetrou:1951:3,Corinaldesi:Papapetrou:1951,Tulczyjew:1959,Tulczyjew:1962,Taub:1964,Dixon:1964,Madore:1969,Dixon:1970:1,Dixon:1970:2,Dixon:1979}. It has also been successfully applied in alternative gravity theories, see \cite{Stoeger:Yasskin:1979,Stoeger:Yasskin:1980,Nomura:Shirafuji:Hayashi:1991} and more recently in \cite{Puetzfeld:Obukhov:2007,Puetzfeld:Obukhov:2008}. Note that reference \cite{Puetzfeld:Obukhov:2007} also contains a short timeline of works. The model under consideration in the present work does {\it not} belong to the very general class of gravitational models analyzed in \cite{Puetzfeld:Obukhov:2007} due to its nonminimal coupling prescription.

\section{Integrated conservation law}\label{integrated_conservation_law_sec}

The multipole scheme employed by us relies on the integration of the conservation law over the world tube of the test body, see also \cite{Puetzfeld:Obukhov:2007} for details. To begin with, we rewrite (\ref{conservation}) as follows:

\begin{eqnarray}
\nabla^i T_{ij} = \left( g_{ij} L_{\rm mat} - T_{ij} \right) \nabla^i A. \label{conservation_expanded}
\end{eqnarray}  
Here we introduced a scalar function $A\left(R\right) := \log\left[ 1 + \lambda f_2 \left(R \right)\right]$. In the following we are going to denote derivatives of this function simply by $A_i := \partial_i A, A_{ij}:=\partial^2_{ij} A$, etc. Raising the indices and rewriting the covariant derivative in (\ref{conservation_expanded}) yields
\begin{eqnarray}
\widetilde{T}^{ij}{}_{,j} &=& \left( \widetilde{\Xi}^{ij} - \widetilde{T}^{ij} \right) A_j  - \Gamma_{jk}{}^i \widetilde{T}^{jk}. \label{conservation_rewritten}
\end{eqnarray}  
In the last equation we introduced the quantity $\Xi^{ij}:=g^{ij} L_{\rm mat}$ as a shortcut. Densities of different quantities are denoted by a tilde ``$\widetilde{\phantom{A}}$''. 
We define the integrated multipole moments of the matter quantities $\widetilde{\Xi}^{ij}$ and $\widetilde{T}^{ij}$ as follows:
\begin{eqnarray}
\overline{T}^{b_{1}\cdots b_{n}ij} &:&=\int \left( \prod\limits_{\alpha =1}^{n}\delta x^{b_{\alpha }}\right) \widetilde{T}^{ij}, \nonumber \\
\overline{\Xi}^{b_{1}\cdots b_{n}ij} &:&=\int \left( \prod\limits_{\alpha =1}^{n}\delta x^{b_{\alpha }}\right) \widetilde{\Xi}^{ij}. \label{definition_moments}
\end{eqnarray}   
Here $\delta x^a:=x^a-Y^a$, and $Y(t)$ parametrizes the world line of the body. Note that here, and in the following, only the last two indices of the integrated quantities are spacetime indices, and the $b_{1} \cdots b_{n}$ label the multipole order of a current. The integrals in (\ref{definition_moments}) are taken over a 3-dimensional slice $\Sigma(t)$, at a time $t$, over the world tube of the test body. We use the condensed notation
\begin{equation}
\int\,f = \int_{\Sigma(t)}\,f(x)\,d^3x.
\end{equation} 
With the definitions in (\ref{definition_moments}) the integrated version of the conservation law (\ref{conservation_expanded}) takes the following general form:
\begin{eqnarray}
&&\frac{d}{dt}\overline{T}^{b_{1}\cdots b_{n}i0} =\sum_{\beta =1}^{n}\left(\overline{T}^{b_{1}\cdots \check{b}_{\beta } \cdots b_{n} i b_{\beta}}-v^{b_{\beta }}{}\overline{T}^{b_{1}\cdots \check{b}_{\beta }\cdots b_{n} i 0}\right) \nonumber \\ 
&&+\int \left( \prod \limits_{\alpha =1}^{n}\delta x^{b_{\alpha }}\right) \left\{  \left( \widetilde{\Xi}^{ij} - \widetilde{T}^{ij} \right) A_j - \Gamma_{jk}{}^i \widetilde{T}^{jk} \right\}, \label{conservation_law_integrated}
\end{eqnarray}
here an inverted circumflex, e.g.\ ``$\check{b}_\beta$'', indicates the omission of an index from a list and $v^a:=dY^a/dt$  

\section{Propagation equations for pole-dipole test bodies}\label{propagation_sec} 

In this section we work out the equations of motion for pole-dipole test bodies. For such bodies only the moments $\overline{T}^{ij},\overline{T}^{aij},\overline{\Xi}^{ij}$, and $\overline{\Xi}^{aij}$ are nonvanishing. With the expansion of geometrical quantities around the worldline $Y(t)$ of the test particle into a power series in $\delta x^{a}$ 
\begin{eqnarray}
\left. R\right| _{x} &=&\left. R\right| _{Y}+\delta x^{a}\left. R_{,a}\right| _{Y}  \nonumber \\
&&+\frac{1}{2} \delta x^{a} \delta x^{a}\left. R_{,ab}\right| _{Y} +\dots ,  \nonumber \\
\left. \Gamma _{ij}{}^{k}\right| _{x} &=&\left. \Gamma _{ij}{}^{k}\right|_{Y}+\delta x^{a}\left. \Gamma _{ij}{}^{k}{}_{,a}\right| _{Y}\nonumber \\
&&+\frac{1}{2}\delta x^{a}\delta x^{b}\left. \Gamma _{ij}{}^{k}{}_{,ab}\right| _{Y}+\dots, 
\end{eqnarray}
the integrated conservation law (\ref{conservation_law_integrated}) yields the following set of propagation equations:
\begin{eqnarray}
\frac{d}{dt}\overline{T}^{i0}&=&-\Gamma_{cd}{}^{i} \overline{T}^{cd} - \Gamma_{cd}{}^{i}{}_{,b} \overline{T}^{bcd} + \left( \overline{\Xi}^{ib} - \overline{T}^{ib} \right) A_c \nonumber \\
&&+ \left( \overline{\Xi}^{cib} - \overline{T}^{cib} \right) A_{bc}, \label{prop_1} \\
\frac{d}{dt}\overline{T}^{ai0}&=&\overline{T}^{ia} - v^a \overline{T}^{i0} - \Gamma_{cd}{}^{i} \overline{T}^{acd} \nonumber \\
&&+ \left( \overline{\Xi}^{aib} - \overline{T}^{aib} \right) A_b, \label{prop_2} \\ 
0 &=&\overline{T}^{jia} - v^a \overline{T}^{ji0} + \overline{T}^{aij} - v^j \overline{T}^{ia0} \label{prop_3}.  
\end{eqnarray}
Here we suppressed the dependencies on the points at which the quantities are evaluated. 

\section{Propagation equations rewritten}\label{rewritten_sec}

To allow for a better comparison with the result in \cite{Papapetrou:1951:3}, we bring (\ref{prop_1}) - (\ref{prop_3}) into a form which closely resembles the form of the covariant equations (5.3) and (5.7) in \cite{Papapetrou:1951:3}.

We start with the following redefinitions of the integrated moments:
\begin{eqnarray}
M^{ab}  := u^0 \overline{T}^{ab}, \quad M^{abc} := - u^0 \overline{T}^{abc}. \label{D1_D2}
\end{eqnarray} 
Here we introduced $u^a := d Y^a / ds$ for the velocity and the parameter $s$ denotes proper time. Furthermore one should note that $M^{0ab}=0$ due to our choice of the integration domain over hypersurfaces with $t={\rm const}$. Analogously to \cite{Papapetrou:1951:3} we introduce the spin as follows:
\begin{eqnarray}
S^{ab}:=\overline{T}^{ab0} - \overline{T}^{ba0}. \label{spin_def}
\end{eqnarray}
From these definitions we can immediately infer that
\begin{eqnarray}
u^0 S^{ab} = - \left( M^{ab0} - M^{ba0}\right). \label{h2}
\end{eqnarray}
With (\ref{D1_D2}) the propagation equation (\ref{prop_3}) becomes
\begin{eqnarray}
u^0 \left(M^{jai} + M^{iaj} \right) = u^i M^{ja0} + u^j M^{ia0}. \label{P3_1}
\end{eqnarray}
Cyclic permutation of the indices in equation (\ref{P3_1}) and subtraction of the second from the sum of the first and the third of the permutations yields
\begin{eqnarray}
\hspace{-0.2cm} 2 M^{iaj} = - \left( u^j S^{ia} + u^a S^{ij} \right) + \frac{u^i}{u^0} \left(S^{0j} u^a + S^{0a} u^j \right). \label{P3_2}
\end{eqnarray}
With the definition 
\begin{eqnarray}
N^{iab}:=u^0 \overline{\Xi}^{iab}, \label{D3}
\end{eqnarray}
the second propagation equation (\ref{prop_2}) becomes
\begin{eqnarray}
M^{ai} &=& u^i \frac{M^{a0}}{u^0} - \frac{d}{ds} \left(\frac{M^{ia0}}{u^0} \right) - \Gamma_{cd}{}^a  M^{icd}\nonumber \\
&& - \left( N^{iab} + M^{iab} \right) A_b. \label{P2_1}
\end{eqnarray}
If we introduce the following ``generalized momentum''
\begin{eqnarray}
\mu^a:= \frac{1}{u^0} \left( M^{a0} + \Gamma_{cd}{}^a u^d S^{c0} \right),\label{def_momentum}
\end{eqnarray}
equation (\ref{P2_1}) can be recast into
\begin{eqnarray}
M^{ai} &=& u^i \mu^a + \frac{1}{2} \frac{DS^{ia}}{Ds} + \frac{d}{ds} \left( \frac{u^{(i} S^{a)0}}{u^0} \right) +u^d \Gamma_{cd}{}^{(a} S^{i)c} \nonumber \\
&& - \left( N^{iab} + M^{iab} \right) A_b.  \label{h5}
\end{eqnarray}
Here we used the definition of the covariant derivative
\begin{eqnarray}
\frac{DS^{ab}}{Ds}:=\frac{dS^{ab}}{ds} + \Gamma_{cd}{}^{a} S^{cb} u^d + \Gamma_{cd}{}^{b} S^{ac} u^d. \label{D5}
\end{eqnarray}
Taking the skew symmetric part of (\ref{h5}) yields
\begin{eqnarray}
u^{[i} \mu^{a]} + \frac{1}{2} \frac{\widehat{D} S^{ia}}{Ds}=0, \label{skew_h5}
\end{eqnarray}
and the derivative with the hat is defined as follows:
\begin{eqnarray}
\hspace{-0.2cm} \frac{\widehat{D}S^{ia}}{Ds} := \frac{D S^{ia}}{Ds} + \left( N^{aib} - N^{iab} + M^{aib} - M^{iab} \right) A_b. \label{D5_extended}
\end{eqnarray}
If we contract (\ref{skew_h5}) with $u_i$ and make the same choice for the mass as in \cite{Papapetrou:1951:3}, namely
\begin{eqnarray}
m:=\mu^i u_i, \label{D6}
\end{eqnarray}
we obtain
\begin{eqnarray}
\mu^a =m u^a + u_i \frac{\widehat{D}S^{ai}}{Ds}. \label{skew_h5_conseq}
\end{eqnarray}
Substituting this equation back into (\ref{skew_h5}) yields a very compact version of the second propagation equation, namely
\begin{eqnarray}
\frac{\widehat{D}S^{ia}}{Ds} + u^i u_c \frac{\widehat{D}S^{ac}}{Ds} - u^a u_c \frac{\widehat{D}S^{ic}}{Ds} = 0, \label{E2_1}
\end{eqnarray} 
which should be compared to (5.3) in \cite{Papapetrou:1951:3}.

Finally, with the definition 
\begin{eqnarray}
N^{ab}:=u^0 \overline{\Xi}^{ab},
\end{eqnarray}
the first propagation equation (\ref{prop_1}) becomes 
\begin{eqnarray}
&&\frac{d}{ds} \left(\frac{M^{a0}}{u^0} \right) + \Gamma_{cd}{}^{a} M^{cd} - \Gamma_{cd}{}^{a}{}_{,b} M^{bcd} \nonumber \\
&& = \left( N^{ab} - M^{ab} \right) A_b + \left(N^{cab} + M^{cab} \right) A_{bc}. \label{P1_1} 
\end{eqnarray}
With the help of (\ref{h5}) and (\ref{P3_2}), and by using the derivatives as defined in (\ref{D5}) and (\ref{D5_extended}) we can bring (\ref{P1_1}) into its final form
\begin{eqnarray}
&&\hspace{-0.5cm}\frac{D}{Ds}\left[mu^a + u_c \frac{\widehat{D}S^{ac}}{Ds} \right] + \frac{1}{2} S^{bc} u^d  R_{bcd}{}^a = \left(N^{ab} - M^{ab}\right) A_b \nonumber \\
&& + \left(N^{cdb} + M^{cdb} \right) \left( \delta^a_d A_{bc} + \Gamma_{cd}{}^a A_b\right).  \label{E1_2} 
\end{eqnarray}
This result should be compared to equation (5.7) in \cite{Papapetrou:1951:3}.

\section{Physical consequences}\label{consequences_sec} 

Our main result is embodied in the propagation equations as given in (\ref{E2_1}) and (\ref{E1_2}). These equations should be compared to the well-known ones for pole-dipole test particles in general relativity as given in (5.3) and (5.7) of \cite{Papapetrou:1951:3}, nowadays usually called the Mathisson-Papapetrou-Dixon (MPD) equations. 

Our result in the case of the nonstandard gravity theory clearly shows the additional terms which arise due to the nonminimal coupling. The structure of the generalized propagation equation for the spin, i.e.\ equation (\ref{E2_1}), is very similar to the classic result in (5.3) of \cite{Papapetrou:1951:3}. In (\ref{E2_1}) the new derivative from (\ref{D5_extended}) takes into account the extra terms which arise due to the nonminimal coupling prescription on the Lagrangian level. The second of the MPD equations can be easily recovered from (\ref{E2_1}) by replacing the ``hatted'' derivative by the standard one given in  (\ref{D5}). Also the first generalized propagation (\ref{E1_2}) allows for a quick recovery of the first MPD equation for the momentum. As soon as one switches off the nonminimal coupling, the rhs of (\ref{E1_2}) vanishes, and the hatted derivative on the left-hand side (lhs) is replaced by the standard derivative. 

It is very interesting to note that, even when we confine ourselves to nonspinning particles -- i.e.\ test particles for which $M^{abc}$ vanishes -- the rhs of (\ref{E1_2}) is nontrivial and the generalized momentum $\mu^i$, which enters on the lhs of (\ref{E1_2}), is given by
\begin{eqnarray}
\mu^i = m u^i + \left(N^{cib} - N^{icb} \right) u_c A_b,
\end{eqnarray}
yielding $\mu^i \neq m u^i$.

Furthermore, if we do not allow for any kind of dipole contribution, neither via $N^{abc}$ nor $M^{abc}$ -- i.e.\ we consider only pole particles, the first propagation equation becomes 
\begin{eqnarray}
\frac{D}{Ds} \left(m u^i \right) = \left(N^{ib} - M^{ib} \right) A_b. \label{single_pole_equation}
\end{eqnarray}
Hence, we still have a nonstandard contribution on the rhs of the first equation of motion, with a direct coupling between the monopole moments of the matter currents and the background geometry as described by the derivative of the function $A(R)$. 

In other words, even single-pole test particles do {\it not} move along geodesics in the theory under consideration. Of course this is in contrast to the standard result in the theory of general relativity.
 
One should note that in the case of single-pole particles -- and by using (\ref{h5}) and (\ref{skew_h5_conseq}) -- equation (\ref{single_pole_equation}) can be brought into the following form:
\begin{eqnarray}
\frac{D}{Ds} \left[ mu^i \left(1+ \lambda f_2 \right) \right] = N^{ib} \left(1+\lambda f_2 \right)_{,b}. \label{single_pole_equation_rewritten} 
\end{eqnarray}
From this equation one can immediately read off the additional contribution due to the nonminimal coupling terms as embodied by $f_2(R)$. On the lhs of the single-pole equation of motion, these terms induce an ``effective mass''. On the rhs of (\ref{single_pole_equation_rewritten}), the nonminimal coupling procedure -- remembering $N^{ab} \cong g^{ab} N$ --  leads to some ``effective pressure'' term.

Equation (\ref{single_pole_equation_rewritten}) provides an interesting interpretation of the aforementioned Pioneer anomaly \cite{Anderson:etal:2002} as a result of the nonminimal coupling of matter and gravity. From (\ref{single_pole_equation_rewritten}) it is obvious that no extra forces act on a test body for gravitational field configurations with vanishing curvature scalar $R =0$.

However, this picture changes as soon as one considers spacetimes with nonvanishing curvature scalar, for example as encountered in cosmological solutions. Albeit not an exact solution in the context of the theory under consideration, the well-known Friedmann-Lema$\hat{\i}$tre-Robertson-Walker (FLRW) metric may serve as an illustrating example. For a linear choice of the function $f_2$ -- or, equivalently, for the first term in a series expansion of $f_2$ -- the extra force entering (\ref{single_pole_equation_rewritten}) would be directly proportional to the curvature scalar. Hence, in a FLRW background one would obtain an additional contribution due to the nonminimal coupling prescription which is proportional to
\begin{eqnarray}
f_2=R_{\rm FLRW} = 6 \left( H^2 (1-q) + \frac{k}{a^2} \right). \label{ricci_flrw}
\end{eqnarray}
Here we made use of the standard definition of the Hubble rate $H:=\dot{a} / a$ and the deceleration parameter $q:= - a \ddot{a} / \dot{a}^2$. The scale factor is a function of time only, i.e.\ $a=a(t)$, and the constant $k$ determines the spatial curvature in the FLRW metric.

Hence, if we consider the motion of a test body, e.g. of the Pioneer spacecraft, taking into account the cosmological expansion of the spacetime background, then one would obtain a correction to its acceleration -- compared to geodesic motion of general relativity -- which is proportional to the Hubble rate $H$ and its derivatives (which can probably be neglected). This is an interesting result which qualitatively agrees with the measured anomalous acceleration of the Pioneer spacecraft which is approximately $cH_0 \sim 7\times 10^{-10}$ ms$^{-2}$. A quantitative analysis could further provide estimates for the nonminimal coupling constant $\lambda$ and for the form of the function $f_2$.

\section{Conclusion}\label{conclusions_sec} 

We worked out the explicit form of the equations of motion for a gravitational theory (\ref{ansatz_lagrangian}) with nonminimal coupling. Our results extend previous works on the equations of motion of such theories and are independent of the specific form of the energy-momentum tensor. 

The coupling between geometric and matter quantities becomes apparent in our framework and should be taken into account in the systematic testing of the theory. Furthermore, our analysis confirms -- in a very general way -- the nongeodesic motion of single-pole test bodies. 

The analysis in this work also applies to several other models with nonminimal coupling. A direct comparison can be made by a simple remapping of the function $A(R)$ in the present analysis. Our results also allow for a straightforward comparison with the equations of motion in other alternative gravity theories \cite{Puetzfeld:Obukhov:2007,Puetzfeld:Obukhov:2008,Stoeger:Yasskin:1979,Stoeger:Yasskin:1980,Nomura:Shirafuji:Hayashi:1991}, which do not belong to the class of $f(R)$ gravity theories.

Furthermore, we have shown that within a nonminimal coupling scheme one can expect an additional acceleration of bodies due to the global influence of an expanding universe. This is particularly interesting in the context of the observed anomalous acceleration of the Pioneer spacecraft, since our results indicate that the ``cosmological order'' of this effect is more than a mere coincidence, but could be ascribed to the nonminimal coupling. 

It would be interesting to carry out a systematic study of the motion of test bodies in specific background spacetimes. Such an analysis should lead to very tight constraints of the free parameters of the nonminimal coupling model. 

\begin{acknowledgments}
The constant support and constructive criticism by F.W.\ Hehl (Univ.\ Cologne) is gratefully acknowledged. DP was supported by the Deutsche Forschungsgemeinschaft (Bonn) under the grant SFB/TR 7. AEI publication number 2008 - 087.
\end{acknowledgments}

\bibliography{micro_bibliography}

\end{document}